# Plasmonic Complex Fluids of Nematiclike and Helicoidal Self-Assemblies of Gold Nanorods with a Negative Order Parameter

Qingkun Liu,[1,2] Bohdan Senyuk,[1] Jianwei Tang,[2] Taewoo Lee,[1] Jun Qian,[2] Sailing He,[2,3] and Ivan I. Smalyukh[1,4,*]

[1]Department of Physics, Material Science and Engineering Program, Department of Electrical, Computer, & Energy Engineering, and Liquid Crystal Materials Research Center, University of Colorado, Boulder, Colorado 80309, USA

[2]Centre for Optical and Electromagnetic Research, Zhejiang University, Hangzhou 310058, People's Republic of China

[3]Department of Electromagnetic Engineering, Royal Institute of Technology, S-100 44 Stockholm, Sweden

[4]Renewable and Sustainable Energy Institute, National Renewable Energy Laboratory and University of Colorado, Boulder, Colorado 80309, USA

*Email: ivan.smalyukh@colorado.edu

**Abstract**

We describe a soft matter system of self-organized oblate micelles and plasmonic gold nanorods that exhibit a negative orientational order parameter. Because of anisotropic surface anchoring interactions, colloidal gold nanorods tend to align perpendicular to the director describing the average orientation of normals to the discoidal micelles. Helicoidal structures of highly concentrated nanorods with a negative order parameter are realized by adding a chiral additive and are further controlled by means of confinement and mechanical stress. Polarization-sensitive absorption, scattering, and two-photon luminescence are used to characterize orientations and spatial distributions of nanorods. Self-alignment and effective-medium optical properties of these hybrid inorganic-organic complex fluids match predictions of a simple model based on anisotropic surface anchoring interactions of nanorods with the structured host medium.

Self-assembly of nanoscale building blocks into long-range three-dimensional (3D) structures is key to the development of nanostructured materials with preengineered optical properties [1–4]. Liquid crystals (LCs) combine fluidity with long-range orientational ordering and thus are promising host media for organizing micro- and nanosized colloids into long-range ordered structures [5–8]. Although a number of different physical mechanisms can be utilized, ranging from LC elasticity to nanoscale segregation, the main idea in these applications of LCs as "smart fluid hosts" for nanoparticle assembly is that they impose various degrees of orientational or positional ordering and serve as bulk fluid templates for colloidal self-organization [8]. Even when dispersed in regular solvents, highly anisotropic rod- and disk-shaped particles at high concentrations can spontaneously organize themselves into LC phases to maximize their translational entropy [5,9]. However, the design and fabrication of nanostructured composites with preengineered properties [10] often require nanoscale organization of weakly anisotropic nanoparticles incapable of forming LC phases and types of ordering inaccessible by maximization of entropy. One of the grand challenges is to construct long-range-ordered fluid composites made from anisotropic nanoparticles forming structures not achievable by means of entropy maximization and elastic interactions in LCs.

In this Letter, we describe soft matter systems with unexpected long-range nematiclike and helicoidal self-assembly of rodlike nanoparticles with a negative orientational order parameter in a discotic LC. These types of colloidal organization have never been achieved before and open new perspectives for the fabrication of reconfigurable composites. Ordering of gold nanorods (GNRs) is probed by polarization-sensitive absorption, scattering, and two-photon luminescence (TPL) and described by a simple model based on anisotropic surface anchoring interactions of nanorods with the LC host. The self-alignment of GNRs in the fluid host results in polarization-sensitive surface plasmon resonance (SPR) effects. Aligned nanoparticles also give rise to dichroism and change of the sign of enhanced optical birefringence within the visible spectral range, potentially enabling applications in tunable polarizing optics.

GNRs with a mean width of 20 nm and a mean length of 50 nm were synthesized using a seed-mediated method [11]. Thiol-terminated methoxy-poly(ethylene glycol) (mPEG-SH, JemKem Technology) was used to coat GNRs for colloidal stability [12]. First, this GNR dispersion was centrifuged at 9000 rpm for 20 min and then resuspended to 1 mL of deionized water to reduce the concentration of cetyltrimethylammonium bromide used for the growth and stabilization of nanorods. Then, 250 μL of an aqueous solution with 2 mM of 5 kDa mPEG-SH was added into 5 mL of a 3.5 nM GNR dispersion. The mixture sat overnight and was purified via centrifugation to eliminate the excess mPEG-SH. We used a ternary lyotropic LC of sodium decyl sulfate-decanol-water (SDS/1-decanol/water) with a known phase diagram [13] for GNR dispersion. Typically, a discotic nematic (ND) dispersion was obtained using a composition of 35.5 wt% of SDS, 7.5 wt% of 1-decanol (both from Sigma-Aldrich), and 57 wt% of an aqueous suspension of mPEG-GNRs at $3\times10^{-7}$ M. This composition was centrifuged at 3000 rpm for 10 min and ultrasonicated for 30 min at room temperature to yield $N_D$ with disk-shaped micelles of about 5.7 nm in diameter and 2.0 nm in thickness. Alignment of a far-field director $\mathbf{N}_0$ was defined by a unidirectional shear and characterized by both polarizing optical microscopy (POM) and two-photon excitation fluorescence polarizing microscopy [14]. The discotic cholesteric ($Ch_D$) dispersion with an equilibrium pitch p ≈ 50 μm was obtained by adding 2 wt% of a chiral agent brucine sulfate heptahydrate (Sigma-Aldrich) to the $N_D$ matrix (37.2 wt% of SDS, 6.2 wt% of 1-decanol, and 54.6 wt% of an aqueous suspension of mPEG-GNRs at $3\times10^{-8} – 3\times10^{-7}$ M) [15]. GNR concentrations are tested using the molar extinction coefficient at the longitudinal SPR peak and the measured optical density. The final concentration of GNRs was about 6% by weight (~0.3% by volume) in $N_D$ and varied within 0.6–6% by weight in $Ch_D$. Long-term stable dispersions were obtained for GNR concentrations up to $3\times10^{-7}$ M and number densities up to 200 $\mu m^{-3}$.

Extinction spectra were obtained using a spectrometer (USB2000-FLG, Ocean Optics) mounted on an optical microscope. POM of GNR-LC composites was performed using 10×, 20×, and 50× dry objectives (all from Olympus) with numerical aperture NA = 0.3–0.9 and a CCD camera (Spot 14.2 Color Mosaic, Diagnostic Instruments, Inc.). Dark-field microscopy was performed using the same microscope equipped with an oil immersion dark-field condenser (NA = 1.2) and a polarizer in the optical path before the camera; only highly scattered light was collected using a 50× air objective (NA = 0.5). TPL imaging [16] was performed using linearly polarized excitation by an 800 nm light from a tunable Ti:Sapphire oscillator (140 fs, 80 MHz, Chameleon Ultra-II, Coherent) at an average power <1 mW in the sample plane [17,18]. A 60× oil objective with NA = 1.42 was used for epidetection of TPL within a 400–700 nm range by a photomultiplier tube (H5784-20, Hamamatsu).

GNRs have longitudinal and transverse SPR modes polarized parallel to their long and short axes, respectively. The longitudinal SPR peak of the extinction spectrum of GNRs in LC is centered at 670 nm and redshifted by 20 nm without any further broadening, as compared to that of GNRs in the growth solution, confirming aggregation-free dispersion in a medium with a higher refractive index. The polarization-dependent excitation of SPR is used to probe colloidal dispersions, yielding a model of self-aligned GNRs in $N_D$ based on polarized extinction spectra and POM, TPL, and dark-field images. GNRs orient, on average, in the plane containing oblate micelles and perpendicular to a director **N** describing the local average orientation of normals to disk-shaped micelles [Figs. 1(a)–1(c)]; GNRs freely rotate around **N**. Microscopy observations (Fig. 1) and spectral characterization show no signs of aggregation of GNRs, as samples are kept for a long time (several months), provided that water evaporation is prevented. Schlieren textures obtained with POM contain half-integer disclinations [Fig. 1(d)] and match dark-field images of scattering patterns obtained for the two orthogonal polarizations shown in Figs. 1(e) and 1(f). The scattered light intensity from GNRs in $N_D$ is the strongest when a polarizer **P** is normal to **N(r)** and minimal for **P** || **N(r)**, confirming that the average nanorod orientation is perpendicular to **N**. Polarization-sensitive TPL imaging with intrinsic 3D resolution [16,19] is used to characterize the orientation and spatial distribution of GNRs. The strongest TPL due to the SPR-enhanced absorption takes place when the wavelength of the excitation light matches or is in the vicinity of the longitudinal SPR peak. TPL intensity is $\propto \cos^4\psi$, where $\psi$ is an angle between the long axes of GNRs and a polarization $\mathbf{P_{ex}}$ of the excitation laser light [16,19], yielding maximum signal when $\mathbf{P_{ex}}$ is perpendicular to a spatially varying **N(r)**. TPL textures of the GNR-$N_D$ composite obtained for two orthogonal $\mathbf{P_{ex}}$ are mutually complementary [Figs. 1(g)–1(i)] and consistent with a POM texture [inset of Fig. 1(g)].

Helicoidal structures of GNRs are obtained in Ch$_D$ with **N** twisted around a helical axis **χ**. The orientation of GNRs is again azimuthally random in the plane of discoidal micelles perpendicular to **N** [Fig. 2(a)], in agreement with the dark-field images that show periodic bright and dark stripes of scattering from GNRs when **P** ⊥ **χ** [Fig. 2(b)] but become uniformly bright at **P** || **χ** [Fig. 2(c)]. 3D patterns of GNR orientations in Ch$_D$ are probed by polarized TPL imaging, revealing that uniformly distributed GNRs self-align with their long axes perpendicular to the helicoidal **N** in the planar cholesteric LC structures with and without various dislocations (Fig. 3). TPL textures taken for orthogonal $\mathbf{P_{ex}}$ are mutually complementary [Figs. 3(a)–3(c)] and reveal twisted configurations of GNRs, such as the one schematically shown in Fig. 3(d). Helicoidal structures are obtained at different GNR concentrations, e.g., at about $10^{-8}$ M [Fig. 3(e)] and at a tenfold higher concentration of $10^{-7}$ M [Fig. 3(f)]. TPL textures reveal the GNR alignment pattern and $\mathbf{N}(\mathbf{r})$ with nonsingular $\lambda^{+1/2}$ and $\lambda^{-1/2}$ disclinations in a snake-shaped undulating cholesteric structure formed in a mechanically deformed $Ch_D$ [Figs. 3(g) and 3(h)].

SPR of aligned nanorods [20] gives rise to anisotropic properties not present in isotropic GNR dispersions and pure LCs. We measure transmittances $T_{\parallel}$ at $\mathbf{P} \parallel \mathbf{N}_0$ and $T_{\perp}$ at $\mathbf{P} \perp \mathbf{N}_0$ to obtain polarized extinction coefficients $\alpha_{\parallel,\perp}^{\text{ext}} = \alpha_{\parallel,\perp}^{\text{abs}} + \alpha_{\parallel,\perp}^{\text{scat}} = -\ln T_{\parallel,\perp}/d$ [Fig. 4(a)], where $d$ is the sample thickness and $\alpha_{\parallel,\perp}^{\text{abs}}$ and $\alpha_{\parallel,\perp}^{\text{scat}}$ are absorption and scattering coefficients, respectively. These coefficients are dependent on a scalar order parameter [21] $S_{\text{GNR}} = \langle P_2(\cos\theta) \rangle$, where $P_2(x)$ is the second Legendre polynomial and $\theta$ is an angle between a nanorod and $\mathbf{N}$. Since the longitudinal SPR is polarized along GNRs, $S_{\text{GNR}}$ is equal to the scalar order parameter of the longitudinal transition dipole moment and can be estimated as $S_{\text{GNR}} = (\alpha_{\parallel}^{\text{ext}} - \alpha_{\perp}^{\text{ext}})/(\alpha_{\parallel}^{\text{ext}} + 2\alpha_{\perp}^{\text{ext}})$ [21]. The presence of longitudinal and transverse SPR peaks at both $\mathbf{P} \parallel \mathbf{N}_0$ and $\mathbf{P} \perp \mathbf{N}_0$ implies that $-0.5 < S_{\text{GNR}} < 1$. By integrating $\alpha_{\parallel}^{\text{ext}}$ and $\alpha_{\perp}^{\text{ext}}$ within 580–800 nm, we obtain $S_{\text{GNR}} = -0.39 \pm 0.01$. Consistently with the TPL and dark-field imaging, this indicates a broad angular distribution with an average nanorod orientation perpendicular to $\mathbf{N}$.

The finite-difference time-domain (FDTD) method is utilized to model the GNR-LC composites by taking material parameters of gold from Ref. [22] and using an average refractive index of the LC $\bar{n}_{\text{LC}} = 1.39$ [13]. Time-dependent Maxwell equations in a differential form are discretized using central-difference approximations in time and space and then solved using the FDTD Solutions software (from Lumerical Solutions, Inc). GNRs are modeled as cylinders capped with two semispheres. To account for the experimental size distribution (determined from transmission electron microscopy images [8]), we average extinction spectra of GNRs with aspect ratios according to this distribution and adjust nanorod concentration so that simulated and experimental extinction spectra match. It is instructive to compare computer-simulated properties at experimental $S_{\text{GNR}} = -0.39$ and at $S_{\text{GNR}} = -0.5$ corresponding to all GNRs aligning perfectly orthogonal to $\mathbf{N}$. For a given value of $S_{\text{GNR}}$, we obtain them as $\alpha_{\parallel} = \tfrac{2}{3}\eta(\sigma_{\parallel} - \sigma_{\perp})S_{\text{GNR}} + \tfrac{1}{3}\eta(\sigma_{\parallel} + 2\sigma_{\perp})$ and $\alpha_{\perp} = -\tfrac{1}{3}\eta(\sigma_{\parallel} - \sigma_{\perp})S_{\text{GNR}} + \tfrac{1}{3}\eta(\sigma_{\parallel} + 2\sigma_{\perp})$ [Fig. 4(b)]

[23,24], where $\eta$ is the number density; $\alpha_{\parallel}$ and $\alpha_{\perp}$ denote extinction, absorption, and scattering coefficients for the two orthogonal polarizations; and $\sigma_{\parallel}$ and $\sigma_{\perp}$ are the corresponding calculated extinction, absorption, and scattering cross sections of the bare longitudinal and transverse SPR modes of GNR. The spectra at $S_{GNR} = -0.5$ obtained in this way are consistent with a direct calculation and averaging of spectra due to GNRs at 40 different azimuthal orientations orthogonal to **N** [Fig. 4(b)]. The longitudinal SPR extinction cross section and molar extinction coefficient are calculated to be $2.0 \times 10^{-11}$ cm$^2$ and $1.2 \times 10^{10}$ M$^{-1}$ cm$^{-1}$, respectively. The scattering coefficient is < 12% of the extinction coefficient. We have extracted the absorption coefficients by matching the experimental and simulated extinction spectra and then subtracting the simulated scattering from the extinction measured experimentally [Figs. 4(a) and 4(b)].

The polarization-dependent imaginary parts of the complex refractive index $\tilde{n} = n + i\kappa$ at **P** || **N**$_0$ and **P** $\perp$ **N**$_0$ are calculated as $\kappa_{\parallel,\perp} = \alpha_{\parallel,\perp}^{abs} \lambda / (4\pi)$ [24] and their anisotropy as $\Delta\kappa = \kappa_{\parallel} - \kappa_{\perp}$ [Figs. 4(c) and 4(d)]. Spectral dispersions of refractive indices $n_{\parallel}$ at **P** || **N**$_0$ and $n_{\perp}$ at **P** $\perp$ **N**$_0$ are obtained using the Kramers-Krönig relation: $n_{\parallel,\perp}(\lambda) = n_{\parallel,\perp}^{offset} + [1/(2\pi^2)] \times$ P.V.$\int_{\lambda_1}^{\lambda_2} \alpha_{\parallel,\perp}^{abs}(\lambda) / [1 - (\lambda'/\lambda)^2] d\lambda'$, where P.V. is the Cauchy principal value of the integral [24], $\lambda_1$ = 450 nm, and $\lambda_2$ = 900 nm. The used values of offset extraordinary and ordinary indices $n_{\parallel,\perp}^{offset}$ are based on $\bar{n}_{LC}$ = 1.39 and the intrinsic optical anisotropy $\Delta n_{LC}$ = 0.006 ± 0.001 of the $N_D$ measured in Ref. [13]. Experimentally and numerically determined spectra of GNR-enhanced optical anisotropy $\Delta n(\lambda)$ of the composite closely match each other and change sign at around the longitudinal SPR peak wavelength [Fig. 4(e)].

Self-alignment of GNRs in $N_D$ and Ch$_D$ can be understood by considering anisotropic surface anchoring interactions. GNRs have dimensions smaller than the anchoring extrapolation length for this system, expected to be within $l_e = K/W$ = 100–1000 nm [25], where $W$ is an anchoring coefficient and $K$ is an average Frank elastic constant. Nanorods produce only very weak elastic distortions of **N**, so that elasticity-mediated colloidal interactions among them can be neglected and have no effects on stability. Colloidal stabilization by mPEG-SH prevents GNR surfaces from approaching each other to distances of about 5 nm and smaller, so that depletion forces between nanorods (due to disk-shaped micelles interspacing them) are negligible, too [26].

The alignment of well-dispersed GNRs arises from anisotropic surface anchoring interactions at LC-nanorod interfaces. Nanoparticles induce vertical boundary conditions for **N**, so that the surface anchoring energy is minimized when GNRs point in directions orthogonal to **N**. Assuming that the polar surface anchoring energy density has the Rapini-Papoular form $f_{sa} = (W/2)\sin^2\beta$ ($\beta$ is an angle between **N** and the "easy" axis normal to the GNR's surface) [27], we find the total anchoring energy as a function of the angle $\theta$ between **N** and the nanorod. The distribution of GNR orientations due to surface anchoring interactions is $f(\theta) \propto \exp(-\delta\cos^2\theta)$, where $\delta = \pi LRW/(2k_BT)$; $L$ and $R$ are GNR's length and radius, respectively; $k_B$ is the Boltzmann constant; and $T$ is temperature. We obtain $S_{GNR} = \int_0^\pi P_2(\cos\theta)f(\theta)\sin\theta d\theta = -3\exp(-\delta)/[2\sqrt{\pi\delta}\,\text{Erf}(\sqrt{\delta})] + 3/(4\delta) - 1/2$ [Fig. 4(f)], where Erf($x$) is an error function. Taking experimental $S_{GNR} = -0.39$, $L \approx 50$ nm, $R \approx 10$ nm, T $\approx$ 300 K, and $\delta \approx 6.8$, we find $W \approx 3.6\times10^{-5}$ J/m$^2$ comparable to that measured at similar LC-solid interfaces [28]. The inset of Fig. 4(f) shows the calculated equilibrium distribution of GNR orientations with respect to **N** at $S_{GNR} = -0.39$. Our simple model explains experimental observations of highly unusual spontaneous ordering of GNRs with a negative scalar order parameter, and it will be of great interest to supplement it with more detailed numerical studies to see how exactly GNR-micelle interactions give rise to this behavior.

To conclude, we have reported the first observation of unexpected nematiclike and helicoidal structured dispersions of gold nanorods with a negative scalar order parameter and both low orientational and low translational entropy. These composites exhibit properties present neither in isotropic GNR dispersions nor in pure LCs, such as polarization-sensitive SPR, large absorption anisotropy, and enhanced optical birefringence with sign reversal at the longitudinal SPR peak. The studied system can be further enriched by additional doping with magnetic nano-needles to enhance the response to magnetic fields [29–31] as well as by adding other anisotropic nanoparticles, ranging from carbon nanotubes to semiconductor nanoprisms [30,32]. Composites of oblate micelles and prolate colloids with large aspect ratios may potentially exhibit biaxial nematic phase behavior. Furthermore, since the capping of GNRs and surfactant micelles can be of both anionic and cationic types (in addition to electrostatically neutral surfactants and polymers), lyotropic LC dispersions of anisotropic nanoparticles with well-defined geometric shapes may exhibit new types of phases. For example, dispersions of specially capped metal and semiconductor nanowires [18,19] may form phases analogous to sliding phases of DNA-cationic lipid complexes [33–35] enriched by well-defined geometric shapes of nanowire cross sections [19,32]. Our findings may also enable the fabrication of plasmonic polarizers that are of great interest for display and other electro-optical applications [36].

This work was supported by the International Institute for Complex Adaptive Matter and by NSF Grants No. DMR-0847782, No. DMR-0820579, and No. DMR-0844115. We acknowledge discussions with Michael Campbell, Noel Clark, Leo Radzihovsky, Yalun Wang, Qiuqiang Zhan, Yuan Zhang, and Pengxin Chen.

**Figures**

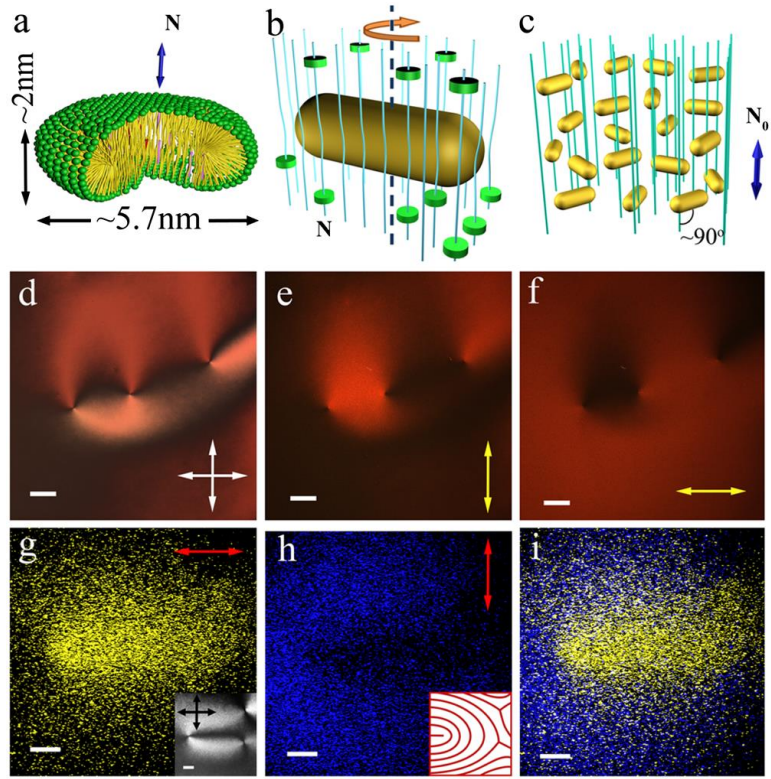

FIG. 1. Schematics of (a) a disk-shaped micelle, (b) a GNR aligned parallel to micelles and perpendicular to **N**, and (c) multiple GNRs self-aligned, on average, perpendicular to **N**$_0$. (d) POM texture of concentrated dispersions of GNRs in $N_D$ obtained between crossed polarizers (white arrows). (e), (f) Polarized dark-field images of the same area as in (d) for two orthogonal polarizations of detected scattering light (yellow arrows). (g),(h) In-plane TPL images of the $N_D$ dispersion for two orthogonal polarizations of excitation light (red arrows). The inset in (g) shows the corresponding transmission-mode micrograph obtained between crossed polarizers (black arrows), and the corresponding **N(r)** is shown in the inset of (h) using red lines. (i) Superposition of the textures of (g) and (h). The scale bars are 20 μm. The intensity of polarized TPL signals in (g) and (h) varies from minimal (black) to maximal (yellow and blue or corresponding gray scale levels in the print version).

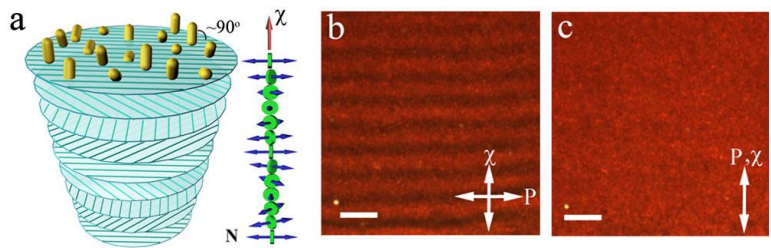

FIG. 2. (a) A schematic of GNRs aligned perpendicular to the helicoidal $\mathbf{N}(\mathbf{r})$ in Ch$_D$. (b),(c) Dark-field images of GNRs at a $3\times10^{-8}$ M concentration for two orthogonal polarizations. The (b) periodic and (c) uniform scattering textures are consistent with the GNR alignment shown in (a). $\mathbf{P}$ and $\chi$ mark the linear polarization and helical axis, respectively. The scale bars are 40 µm.

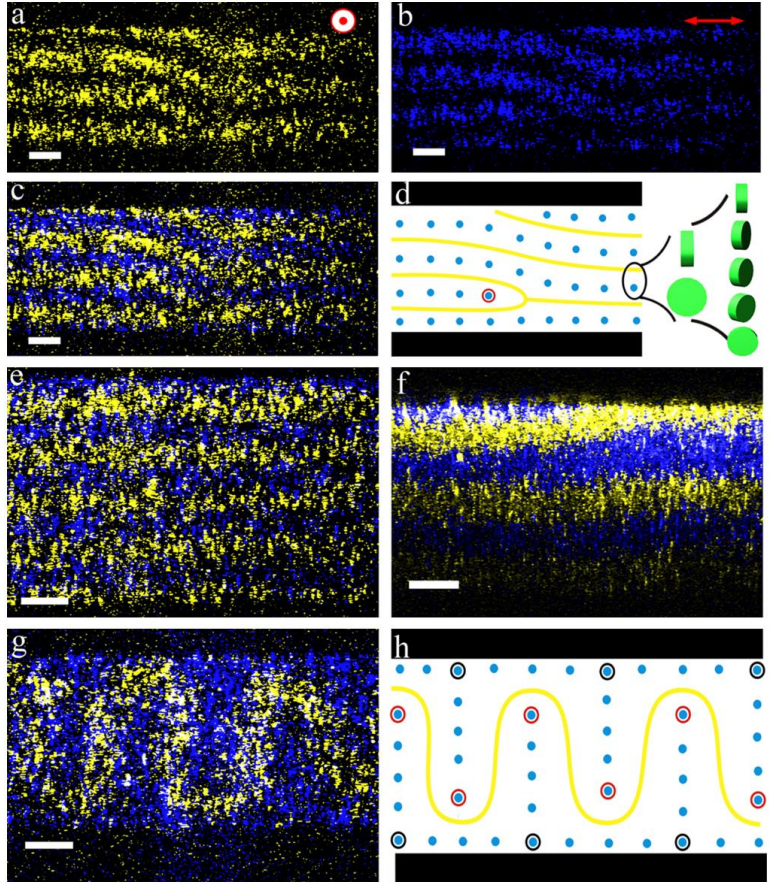

FIG. 3. (a),(b) Vertical TPL cross sections for two orthogonal $\mathbf{P_{ex}}$ for GNRs at a concentration of $3\times10^{-8}$ M in the Ch$_D$ layered structure with a dislocation. (c) Superposition of (a) and (b). (d) A schematic of $\mathbf{N(r)}$ corresponding to the TPL signal shown in (c). (e),(f) Superimposed TPL images obtained for two orthogonal $\mathbf{P_{ex}}$ in planar Ch$_D$ cells at (e) $3\times10^{-8}$ M and (f) $3\times10^{-7}$ M of GNRs, respectively. (g) TPL image of a snake-shaped pattern in a dilated Ch$_D$ planar cell with an undulating structure and (h) a schematic of the corresponding $\mathbf{N(r)}$. The TPL signal coded by the yellow (light gray) color corresponds to $\mathbf{P_{ex}}$ perpendicular to the image plane [shown by a red double circle in (a)], and the signal depicted by the blue (dark gray) color corresponds to $\mathbf{P_{ex}}$ parallel to a horizontal edge of the image [red arrow in (b)]. The red/gray and black circles in (h) and (d) indicate $\lambda^{+1/2}$ and $\lambda^{-1/2}$ disclinations, respectively. The scale bars are 20 µm. The intensity of polarized TPL signals varies from minimal (black) to maximal (yellow and blue or corresponding gray scale levels in the print version).

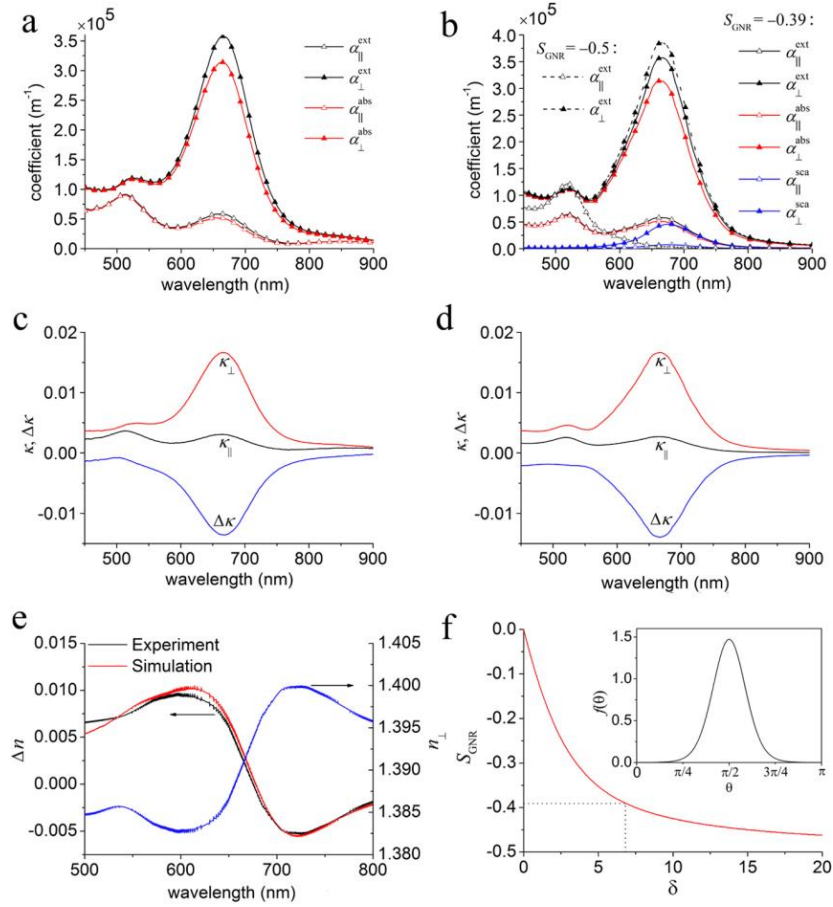

FIG. 4. (a) Experimental extinction and extracted absorption spectra of GNRs in $N_D$ at $3\times10^{-7}$ M. The experimental spectra are measured using unpolarized light first passing through a 10 μm thick cell with aligned GNRs and then through a linear polarizer either parallel or perpendicular to $\mathbf{N}_0$. (b) Simulated extinction, absorption, and scattering coefficients for the same concentration of GNRs in $N_D$ as in (a) for two orthogonal polarizations of incident light at $S_{GNR} = -0.5$ and $S_{GNR} = -0.39$. (c) Experimental and (d) simulated $\kappa_\parallel$, $\kappa_\perp$, and $\Delta\kappa$. (e) Experimental (black line) and computer-simulated (red/gray line) spectral dependence of $\Delta n$ and $n_\perp$ (blue/gray line, experimental). (f) $S_{GNR}$ vs $\delta$. The inset shows the distribution of GNR orientations with respect to $\mathbf{N}$ at $S_{GNR} = -0.39$.